\documentclass[aps,pra,showpacs,twoside,twocolumn,10pt]{revtex4-2}
\usepackage[colorlinks=true, citecolor=blue, urlcolor=blue ]{hyperref}
\usepackage{epsfig,newlfont,amssymb,amsfonts,amsmath,bm,subfigure,palatino,mathtools,amsthm,braket,soul,enumitem,xcolor,graphics,graphicx,times,physics}
\usepackage[normalem]{ulem}

\usepackage[left=20mm, right=20mm, top=0.7in, bottom=0.7in]{geometry}

\usepackage{graphicx}

\usepackage {graphics}
\usepackage {array}

\usepackage[normalem]{ulem}

\makeindex

\begin{document}

\title{Expediting quantum state transfer through the long-range extended  $XY$ model }
\author{Sejal Ahuja$^1$, Tanoy Kanti Konar$^{1,2}$, Leela Ganesh Chandra Lakkaraju$^{1,3,4}$, Aditi Sen(De)$^1$}

\affiliation{$^1$Harish-Chandra Research Institute, A CI of Homi Bhabha National Institute,  Chhatnag Road, Jhunsi, Prayagraj - 211019, India}
\affiliation{$^2$Institute of Theoretical Physics, Faculty of Physics, Astronomy, and Computer Science, Jagiellonian University in Krakow, Stanisława \L{}ojasiewicza street 11, PL-30-348 Krak\'ow, Poland}
\affiliation{$^3$Pitaevskii BEC Center, CNR-INO and Dipartimento di Fisica, Universit\`a di Trento, Via Sommarive 14, Trento, I-38123, Italy}
\affiliation{$^4$ INFN-TIFPA, Trento Institute for Fundamental Physics and Applications, Trento, Italy}

\begin{abstract}

Going beyond short-range interactions, we explore the role of long-range interactions in the extended $XY$ model for transferring quantum states through evolution. In particular, employing a spin-1/2 chain with interactions decaying as a power law, we demonstrate that long-range (LR) interactions significantly enhance the efficiency of a quantum state transfer (QST) protocol, improving the achievable fidelity, mitigating its slow decline as compared with the nearest-neighbor setting, associated with increasing system-size. Our study identifies the LR regime as providing an optimal balance between interaction range and transfer efficiency, outperforming the protocol with the short-range interacting model. Our detailed analysis reveals the impact of system parameters, such as anisotropy, magnetic field strength, and coordination number, on QST dynamics.  Specifically, we find that intermediate coordination numbers lead to a faster and more reliable state transfer, while extreme values diminish performance. Furthermore, we exhibit that the presence of LR interactions considerably reduces the minimum time required to achieve fidelity beyond the classical limit.

\end{abstract}
\maketitle

\section{Introduction}
\label{intro}

Quantum technology presents an exciting avenue for the development of advanced devices, such as communication systems \cite{BBCJPWtele,Pan2000Feb,Kumar2019Sep} including cryptography \cite{buzek_secret_sharing, gisinrmp}, quantum batteries \cite{Alicki, battery_rmp_review}, sensors \cite{degan_review,QmetroRMP}, that could significantly surpass the performance of existing classical counterparts.
One essential component in quantum networks and circuits is the transmission of information through quantum channels, known as quantum state transfer (QST). In this framework, physical systems such as engineered spin chains in condensed matter systems have emerged as effective data buses \cite{Bose03,subrahmanyam_pra_2004,almeida_pra_2018}, facilitating the transmission of quantum information between different nodes which has been shown in different setup \cite{databus_pra_2005,burgarth_1,jian_pra_2007,chen_pra_2007,sanpera_pra_2007,sougata_2,Kay2010Jun,yao_prl_2011,rozhin_pra_2020,Apollaro2022Dec,Jameson2023Nov,Xu2023Nov,david_scipost_2023,huang_prb_2024}. When these spin chains are carefully engineered, they can achieve perfect quantum state transfer under specific conditions \cite{pst_1,pst_2, sougata_1,pst_3,pst_4,pst_5,pst_6,pst_7,pst_8,Pushpan2024Jul}.
This promising approach demonstrates the potential of spin chains as well as several different physical substrates, e.g. photonic systems \cite{anuradha2023,wang_pra_2024} and, superconducting circuits \cite{Liu2023Jun} for quantum state transfer, offering valuable insights for future advancements in quantum communication technology.

\begin{figure}
    \centering
    \includegraphics[width=\linewidth]{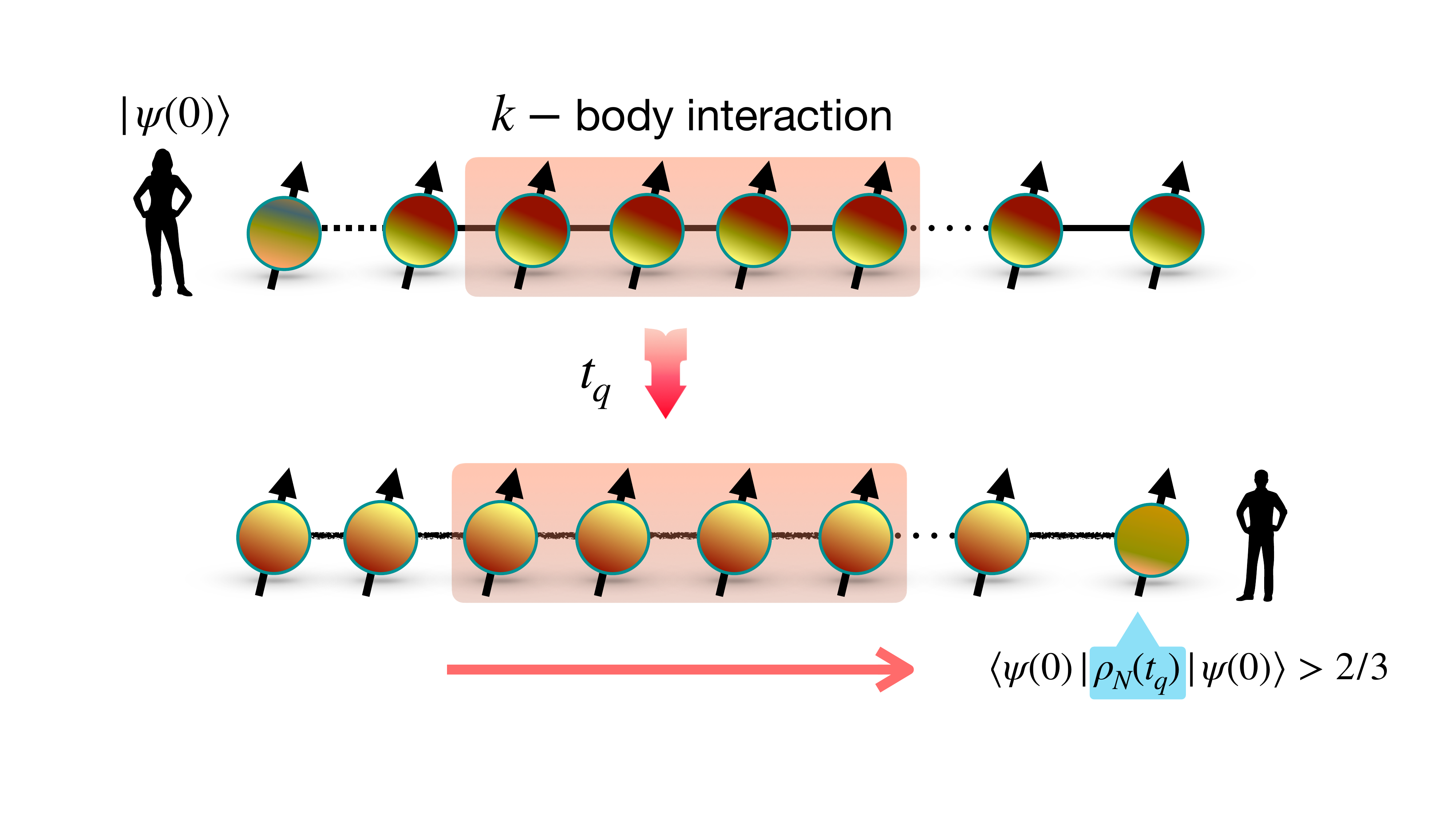}
    \caption{Schematic diagram illustrating quantum state transfer. The first spin in the first line possessed by Alice is the arbitrary state to be transferred while the rest of the sites representing the spin chain involving \(k\)-body interactions acted as a quantum channel. After a time \(t_q\) (second line), the state transfer occurs since the fidelity of the last site crosses the classical limit, \(2/3\). }
    \label{fig:schematics}
\end{figure}

Long-range interacting Hamiltonian  exhibit counter-intuitive phenomena, including many-body localization \cite{mbl_longrange}, alterations in the Lieb-Robinson bound due to the induction of long-range (LR) correlations \cite{lieb_robinson_pra_2020}, and exotic quantum phases and transitions \cite{long_range_review}. These features make long-range systems highly promising for the development of quantum technologies like heat engines, batteries, sensors, and more \cite{heat_engine_longrange,long_range_battery,longrange_sensor,GaneshLRsensing}. Specifically, by investigating correlation length \cite{VodolaLR,GoroshkovLR}, entanglement \cite{entropy_long_range_maciej_luca_prl_2012} and information spreading via Lieb-Robinson bounds \cite{philip_taglia}, it can be found that the system possesses long-range when \(0\leq \alpha \leq 1\), quasi long-range with \(1< \alpha \leq 2\) and short-range for \( \alpha > 2\), \textcolor{black}{where \(\alpha\) is the strength of power-law decay which quantifies how the interaction strength between two sites decays with distance.} Moreover, in trapped ion systems, ions can interact with each other via collective modes of motion, which naturally results in  long-range interactions \cite{bloch05,blochrmp} while  dipolar interactions or the use of optical fields that mediate spin interactions over long distances can be used to engineer cold atoms in optical lattices to display LR interactions \cite{jacpo_prb_2023}. Hence, these LR systems  are not only theoretically intriguing but also experimentally accessible on current platforms.

Even though the potential of long-range interacting systems has been widely recognized through the development of various quantum devices, their application in quantum state transfer remains largely underexplored {\cite{gorshkov_qst_1,gorshkov_LR_4,herms_pra_2020,gorshkov_qst_3,hong_pra_2021}. Notably, specific LR interactions have been  shown to enable perfect QST \cite{pst_3,pst_6}. Motivated by the exciting new paradigm of LR-extended models, we explore their ability to perform efficient QST in realistic scenarios \cite{wang_prr_2024}. Moreover, it is crucial to investigate whether they can outperform systems relying solely on nearest-neighbor (NN) interactions.

To address these challenges, we focus on the extended  $XY$ model \cite{VodolaLR,GoroshkovLR}, involving  \(k\)-body interactions that decay with distance, exhibiting two transitions, from long- to quasi long-range and quasi long- to short-range regimes. Furthermore, it can be solved analytically using Jordan-Wigner, Fourier, and Bogoliubov transformations \cite{lieb1961,barouch_pra_1970_1,barouch_pra_1970_2,sachdev09}, enabling investigations of large system sizes. Importantly, it has been shown to be experimentally realizable on platforms such as optical lattices and trapped-ion systems \cite{long_range_review}. Recently, experiments with more than two-body interactions have been studied in the context of quantum approximate and optimization algorithms \cite{qaoa_review}. Specifically, the three-body $ZZZ$-interaction has been realized in superconducting  \cite{martinis_ZZZ_PRAppl_2024,ZZZ_experiment} and quantum annealing platforms \cite{Pelofske2024Mar}. By leveraging this model, we establish the potential of \(k\)-body interacting LR interactions for robust and efficient quantum state transfer. 

We employ two key figures of merits: the maximum fidelity \(f^*\)above the classical limit (\(f=2/3)\) (see Fig. \ref{fig:schematics}) achieved for the first time and the minimum time required to achieve fidelity beyond the classical limit which we denote as \(t_q\). We identify that the quasi long-range interaction regime offers a distinct advantage over the deep long- and short-range interacting systems. This suggests that there exists an optimal range of long-range interaction strength that is particularly beneficial for QST protocols. We analyze the impact of system-size on the average fidelity of QST. While fidelity decreases with increasing system-size, our findings reveal that in systems with long-range interactions, this reduction is slower. The rate of decline in fidelity depends on the strength of the long-range interactions, further emphasizing the role of long-range effects in mitigating the challenges posed by larger system sizes. Moreover, we exhibit that there is always a coordination number incorporating long-distant interactions that requires less time to surpass the classical fidelity, in comparison to systems with solely NN interactions. \textcolor{black}{In order to understand the physical mechanism behind our state transfer protocol, we analyze the entanglement dynamics by introducing an additional qubit into the system, which is initially entangled maximally with the first qubit  of the chain. We observe that the logarithmic negativity exhibits a sharp peak at the same time when the fidelity achieves its maximum value which provides a physical picture to the QST protocol.} However, our investigations indicate that this quantum advantage is highly dependent on the system parameters, necessitating careful selection of these parameters for optimal performance. 

Overall, our study highlights the critical role of long-range interactions, coordination number, and system parameters in enhancing QST performance. By carefully tuning these factors, it is possible to exploit the exotic properties of the extended $XY$ model to achieve efficient quantum state transfer, even in larger systems. This work provides valuable insights into the practical utility of long-range interactions and offers guidance for designing QST protocols leveraging extended $XY$ models.

The paper is organized in the following manner: Sec. \ref{sec:LRspinmodel} introduces the set-up to achieve the  successful state transmission protocol with the aid of evolving interacting Hamiltonian. In Sec. \ref{sec:fidelity}, we address the question of maximum fidelity that can be attained with the LR spin model during the dynamics and its scaling with the length of the spin chain. Finally, in Sec. \ref{sec:quantadv} we demonstrate how the minimum time to achieve nonclassical fidelity depends on the range of interaction and fall-off rates.   Sec. \ref{sec:conclu} summarizes the results and concluding remarks.

\section{A long-range spin model as a quantum channel for state-transmission}
\label{sec:LRspinmodel}

In a seminal work of quantum teleportation \cite{BBCJPWtele}, it was shown that a sender, say Alice, \(A\) can send an arbitrary quantum state (say, qubit) to the  receiver, say Bob, \(B\) provided they {\it a priori} share an entangled quantum channel. Note that if Alice and Bob do not share any entangled quantum channel, the transmission of an arbitrary qubit cannot be  possible over a fidelity of \(2/3\), which is referred to as classical fidelity \cite{popescu10}. Importantly, this protocol requires entangled measurement at the sender's end while local unitaries at the receiver's side have to be performed after classical communication of Alice's measurement results.  

On the other hand, in the state transfer protocol \cite{Bose03}, the measurement and local unitary is replaced by the evolution of the system by a suitable Hamiltonian, although the goal is same in both the protocols. The dynamical state entangles all the spins, thereby acting as an entangled quantum channel, responsible for the transfer of quantum state from one site to the other.  

Let us set the stage for the protocol. First, prepare an arbitrary state $|\psi(0)\rangle$, which is to be transferred, at the first site of the chain consisting of \(N\)
spin-\(1/2\) particles which interact according to some Hamiltonian \(\hat{\mathcal{H}}\). Initially, the \((N-1)\)-party state is prepared in the ground or canonical equilibrium state of \(\hat{\mathcal{H}}\), i.e., at time \(t=0\),

\begin{eqnarray}
    \hat{\rho}(0) &=& |\psi(0)\rangle \langle \psi(0)|\otimes  \hat{\rho}^{\beta}(0)
    \end{eqnarray}
where  \( \hat{\rho}^{\beta}(0) = e^{-\beta \hat{\mathcal{H}}}/\mathcal{Z}\) with the partition function \(\mathcal{Z} = \text{tr}( e^{-\beta \hat{\mathcal{H}}})\), and \(\beta= 1/k_B T\) having temperature \(T\) and Boltzmann constant \(k_B\), represents the \((N-1)\)-party thermal state of \(\hat{\mathcal{H}}\) acting as a quantum channel for state transmission (see Fig \ref{fig:schematics} for schematic illustration). In our analysis, we consider the thermal state with \(\beta \rightarrow \infty\), i.e., when the system is in a ground state of \( \hat{\mathcal{H}}\). 
Let us now evolve \(N\) sites according to \( \hat{\mathcal{H}}\), resulting in a \(N\)-party state as 
\(|\psi_N(t) \rangle = e^{-i\hat{\mathcal{H}}t} [|\psi(0)\rangle \otimes |\Psi(0)\rangle]\),
where \(|\Psi(0)\rangle\) is the \((N-1)\)-party ground state of the Hamiltonian. At a suitable time, by tracing out \((N-1)\) parties from \(|\psi_N(t) \rangle\),  the resulting state, \(\rho_N(t)\), at site \(N\) is compared with the initial state \(|\psi(0)\rangle\) at the first site by computing the average fidelity for several initial states, which is given as 
\begin{eqnarray}
f &= &\int \langle \psi_U |\, \rho_{N}(t)\, | \psi_U \rangle \, dU \nonumber \\&=&\int \langle \psi(0)| U \, \rho_{N}(t) U^\dagger \,|\, \psi(0) \rangle \, dU \nonumber \\
&=& \frac{1}{4 \pi} \int_{\theta =0}^{\pi} \int_{\phi=0}^{2\pi}  \langle \psi(0)|\rho_N(t)|\psi(0)\rangle \sin \theta d\theta d\phi,
\label{eq:fid}
\end{eqnarray} 
\textcolor{black}{where, in the first and the second lines,  the integral is taken over the entire Bloch sphere, i.e., \(U\) is sampled according to the Haar measure with \(\int dU=1\) and  the third line is written by parametrizing the initial state as \(|\psi(0)\rangle = \cos \frac{\theta}{2} |0\rangle + e^{i \phi }\sin \frac{\theta}{2} |1\rangle\), with $0 \le \theta \le \pi$ and $0 < \phi \le 2\pi$.} A state transfer protocol is called successful when $f > 2/3$ because it is known that, without an entangled channel, state cannot be transferred with  a fidelity greater than $2/3$ \cite{MassarPopescu95}, referred to as the classical fidelity or limit. This criterion distinguishes quantum protocols from their classical counterparts, establishing $f > 2/3$ as a necessary condition for demonstrating the \textit{quantum advantage}.

The entire protocol depends on the initial preparation of the \((N-1)\)-party state, the evolution operator, thereby depending on the Hamiltonian,  time and the length of the chain \(N\).  Typically, the Hamiltonian responsible for the unitary dynamics involves nearest-neighbor interactions \cite{Bose03,Burgarth2007Dec}.

In our work, the Hamiltonian involving  long-range interacting $XY$-model is considered, given by \cite{GoroshkovLR}
\begin{eqnarray}
    \nonumber \hat{\mathcal{H}} &=&\sum_{j=1}^{N} \sum_{k=1}^{z} -\mathcal{J}_k \Bigg[\frac{1+\lambda}{4}\hat{S}_j^x\hat{\mathcal{Z}}_k^z\hat{S}_{j+k}^x\nonumber+\frac{1-\lambda}{4}\hat{S}_j^y\hat{\mathcal{Z}}_k^z\hat{S}_{j+k}^y\Bigg]\nonumber\\
    &-&\frac{g'}{2}\sum_{j=1}^{N}\hat{S}_j^z.
    \label{eq:LRHam}
\end{eqnarray}
Here, $\hat{\mathcal{Z}}_k^z = \prod_{l=j+1}^{j+k-1} \hat{S}_{l}^{z}$ represents the string operator with $\hat{\mathcal{Z}}_1^z = \hat{\mathbb{I}}$, and $\hat{S}^{\nu}$ ($\nu = x,y,z$) are the Pauli matrices. The coupling strength follows a power-law decay, $\mathcal{J}_k = J/k^\alpha$, where $\alpha$ characterizes the decay strength. The coordination number $z$ determines the range of interactions, analogous to the crystal structures in solid-state systems. When $\alpha \to \infty$, the system reduces to nearest-neighbor interactions, while $\alpha = 0$ corresponds to $z$-neighbor interactions. The parameter $\lambda$ controls the anisotropy in the \(xy\)-plane, and $g = g^\prime/J$ represents the strength of the transverse magnetic field. \textcolor{black}{The Hamiltonian includes string operators of all possible lengths up to order \(N\), with the strength of each term decaying polynomially with distance. As discussed in Ref.~\cite{VodolaLR}, it is precisely the combined effect of these power-law interactions and the associated many-body string terms that enable the transfer of an arbitrary quantum state across the chain.} Note that the Kac normalization is not applicable in our context, as we focus mostly on the regime of $\alpha>1$, where the physics described by both “Kac on” and “Kac off” models is similar \cite{long_range_review}. The advantage of using the Kac normalization is to obtain non-divergent observables in the thermodynamic limit. However, in the context of quantum state transfer, the $N\rightarrow \infty$ limit is inconsequential, as it cannot provide a quantum advantage.




The analysis focuses on chains with lengths around \(N \sim 20\) or \(N \sim 30\) which is typical in current experimental realizations of quantum circuits. Since we are interested in a finite-size system and we also observe that \(\alpha \geq  5\) actually mimics the results obtained with the nearest-neighbor Hamiltonian in the \textcolor{black}{literature \cite{Bayat11}, it is plausible that when the fall-off rate belongs to the neighborhood of \(\alpha=2 \), i.e., \( \alpha \in [2+\epsilon, 3-\epsilon']\), where \(\epsilon\) and \(\epsilon'\) are small numbers, the evolving Hamiltonian still carries  signature of the LR model.} \textcolor{black}{Note that the quantum state transfer dynamics are consistent with the constraints imposed by the Lieb–Robinson bound (LRB) \cite{gorshkov_LR_1,gorshkov_LR_2,gorshkov_LR_3,gorshkov_LR_4,long_range_LR_higher_dimension,Eisert_LR_long_range,long_range_LR_2,gorshkov_LR_5}. However, the \(\alpha-\)transition found from the LRB~\cite{philip_taglia},  which deals with the spread of correlation in the thermodynamic limit, may not be directly reflected in the performance of QST protocol. Our results indicate that while \(\alpha > 5\) effectively corresponds to  nearest-neighbor behavior, \(\alpha \ll 5\) retain the signature of LR interactions.}

To make the study systematic, we consider two extreme \(z\) values \textit{viz}, \(z=N-1\), which corresponds to the interactions between any two arbitrary sites of the chain and  \(2 \leq z\leq N/2\), among which \(z=2\) involves nearest and next-nearest neighbor interactions. Note here that  \(z=1\), representing Hamiltonian only with nearest neighbor interactions, does not have any \(\alpha\)-dependence by definition. Hence, for a fixed \(N, \alpha\) and \(z\), the evolving Hamiltonian responsible for a state transfer is now a function of \(\lambda \geq 0\) and \(g\geq 0\) among which the \(XX\) model with \(\lambda =0\)  does not depend on \(g\) \cite{Bayat11}.

\section{Providing high fidelity with variable-range interacting model }
\label{sec:fidelity}

 Consider a situation where the goal is to transfer an arbitrary quantum state over a chain of fixed length, say, \(N\), via dynamics. Due to a finite size, information transmission can occur at certain times, which implies that the fidelity oscillates over time of evolution. From a practical perspective, it is intriguing to ask -- ``{\it can the fidelity reach a higher value for a long-range interacting model compared to short-range ones?}'' In other words, is there any gain in terms of fidelity by increasing the range of interactions? We answer this question affirmatively.

To address this question, we define the quantity \(f^*\) as the first local  maximum of the fidelity exceeding the classical threshold \(2/3\). Hereafter, the term maximum fidelity refers specifically to this first local maximum. Our goal in this section is to study how \(f^*\) depends on the system parameters, especially when the variable-range interacting $XY$ spin model, in Eq. (\ref{eq:LRHam}), is used to evolve the system (see Appendix  \ref{sec:appendixA} for diagonalization method of the model and \ref{sec:appendixBfid} for the computation of fidelity). In particular, we connect \(f^*\) with the coordination number \(z\) and the fall-off rate of the interaction strength, \(\alpha\) along with the magnetic field strength \(g\), anisotropy \(\lambda\), and the system-size, \(N\). To examine the role of coordination number and fall-off rates, we also fix the  set of parameters, \(\{\lambda,g\} = \{0.9,0.7\} \text { and } \{1,1.7\}\). We justify these values also in the succeeding section based on another quantifier relevant for our study. It is also important to note here that there are other parameter regimes as will be clear in the discussion below in which the similar features emerge.



\begin{figure}
    \centering
\includegraphics[width=\linewidth]{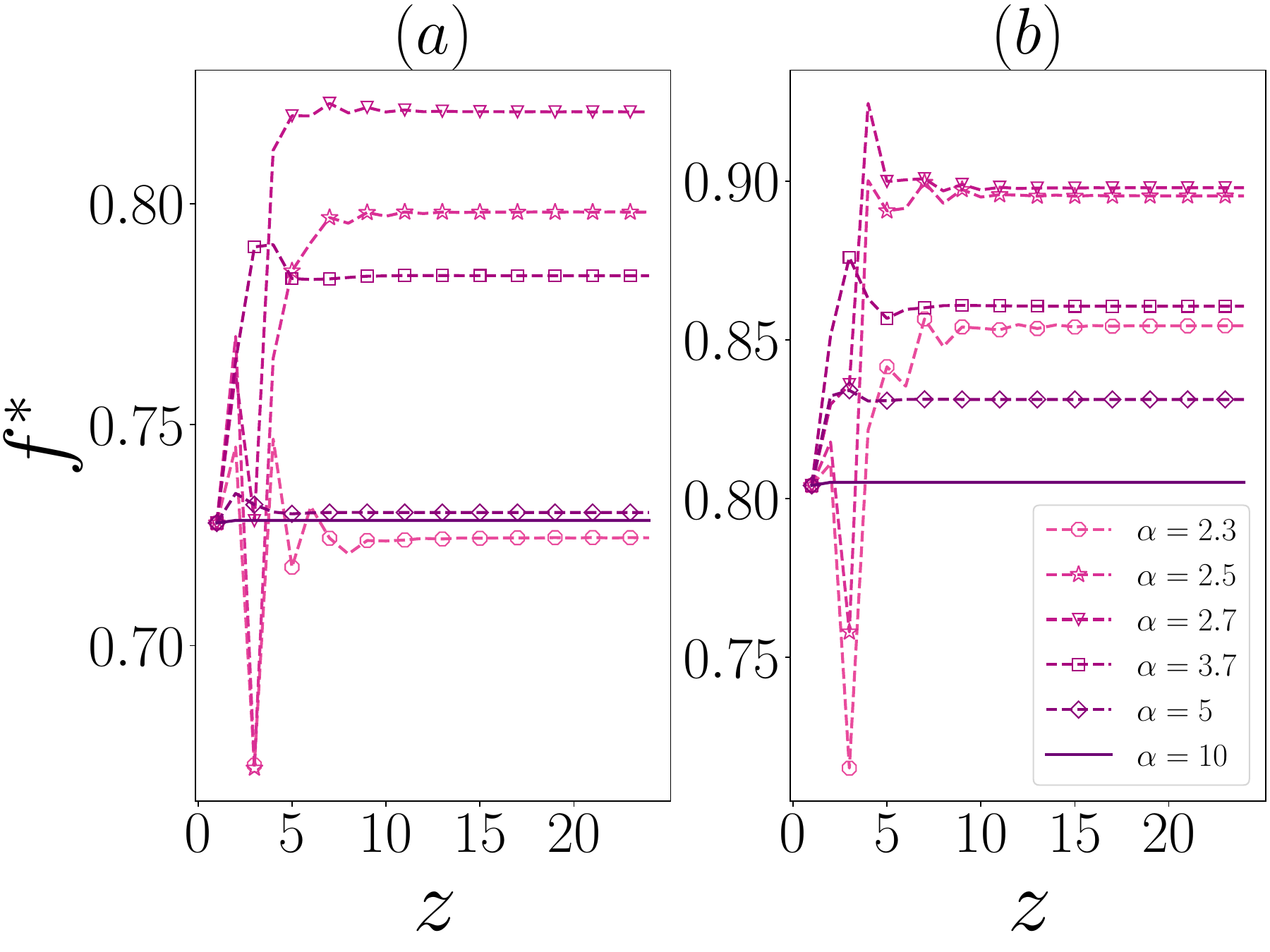}
    \caption{\(f^*\) -- earliest local maximum of fidelity above \(2/3\) $($vertical axis$)$ as a function of the coordination number \(z\) $($horizontal axis$)$ for different fall-off rates.  From the light  to dark  shades, \(\alpha\) increases. Here \(N = 25\).  Two combinations of \( (\lambda, g) \) are considered -- (a) \(\lambda = 0.9\) and \(g = 0.7\), and  (b)  \(\lambda = 1\) and \(g = 1.7\). \(\alpha \geq 5\) mimics \(f^*\) with NN interacting evolving Ising Hamiltonian. Hence, low \(\alpha\) incorporating distant range interactions can yield higher fidelity as compared to the short-range model subject to the choices of \((\lambda, g)\) values. All the axes are dimensionless. }
    \label{fig:fstartwithzdiffalpha}
\end{figure}

{\it Dependence of coordination number.} For a given \(N\), \(g\) and \(\lambda\), we first find \(\alpha\) for which \(f^*\) does not vary with \(z\). We observe that \(f^*\) is independent of \(z\)  when \(\alpha \geq 5\). This suggests that the range of interactions and fall-off rate have a significant contribution in \(f^*\) when \(\alpha < 5\).   The maximum fidelity \(f^*\) has a universal feature with \(z\) for \(\alpha > 2\): \(f^*\)  sharply {changes} with \(z\) and saturates to a value, referred to as \(f^{{*}^{sat}}_z\). 

\begin{table}[h!]
\centering
\begin{ruledtabular}
\begin{tabular}{ccccc}
$\lambda$ & $g$ & $\alpha$ & $z$ & $f^*$ \\
\hline
1   & 1.7 & 0.4 & 2  & 0.82 \\
0.9 & 0.7 & 1.2 & 2  & 0.75 \\
1   & 1.7 & 2.7 & 4  & 0.92 \\
0.1 & 0.1 & 2.6 & 8  & 0.72 \\
0   & 0.7 & 0.6 & 16 & 0.76 \\
0.7 & 1.7 & 3.6 & 24 & 0.87 \\
\end{tabular}
\end{ruledtabular}
 \caption{The dependency of \(f^*\) -- earliest local maximum of fidelity above \(2/3\) with various system parameters. System size \(N\) is fixed at \(25\). It is evident that the presented state transfer protocol has the potential to provide optimal performance subject to the selection of suitable parameters. Note that since the system-size considered here is finite and moderate, the transition points of \(\alpha\) known in the literature may not be valid here, and all \(\alpha\) values chosen in the table possibly carry signature of LR interactions.  }
\label{table:1}
\end{table}

\begin{figure}
    \includegraphics[width=\linewidth]{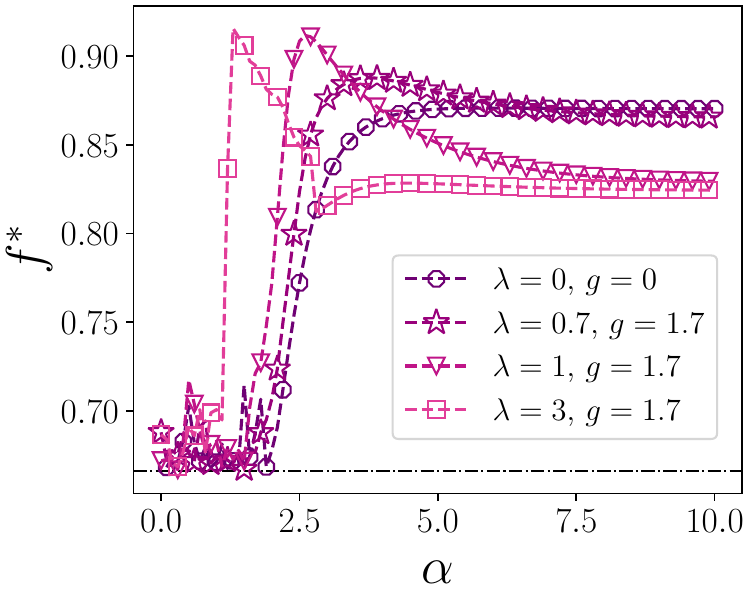}
    \caption{Nonmonotonic behavior of \(f^*\) $($ordinate$)$ with respect to decay strength \(\alpha\) $($abscissa$)$ of the extended transverse Ising model. Different symbols represent choices of \(\lambda\) and \(g\).  Here \(N = 20\) and $z = N-1$. It is observed that the fidelity achieves a maximum value $($even beyond 0.9$)$ for a specific value of \(\alpha\ \sim 2\). Again for small \(\alpha\) values, \(f^*\) fluctuates. All the axes are dimensionless. }
    \label{fig:fstarwithalpha}
\end{figure}

From the study, it is clear that \(z=2\) along with \(z=3\) turn out to be special (see Fig. \ref{fig:fstartwithzdiffalpha}). Since we aim to highlight the benefit of the LR interacting system, we always compare the result with \(z=1\). In particular, we find several instances when \(\alpha \in [2, 4]\), where 
\begin{eqnarray}
    f^{*^{z=1}} <  f^{*^{z=2}}, \, \, \text{and}\,\,   f^{*^{z=1}} <  f^{*^{sat}}_z.
\end{eqnarray}
 In some cases, the increment is significant. For example, we notice that \(\lambda =1\),  \(g=1.7\), \(N =25\),   with \(\alpha =10\), \( f^{*^{sat}}_z =0.81\)  while for the same set of values and  for \(\alpha = 2.7\),  we obtain \( f^{*^{sat}}_z =0.9\). Such a rise of \(f^*\) is also achieved when \(\lambda \neq 0\) and \(|g|<1\), as shown in Fig. \ref{fig:fstartwithzdiffalpha}.  This behavior can be explained in the following way -- a successful execution of the state transfer scheme depends both on the creation of entanglement between different sites, including nearest-neighbor, and at the same time, it is also essential to disentangle the site on which the state is to be transferred. Hence, this could be associated with the competition between entangling and disentangling power of the unitary responsible for transferring the state efficiently, which is lowest when \(\alpha \sim 2 \) and, therefore, it still carries the effect of long-range interactions.

{\it Fidelity with maximum coordination number -- status of anisotropy. } To scrutinize the consequence of long-range interactions, let us consider the maximum coordination number for a fixed \(N\), i.e., \(z=N-1\). The observations are as follows: 
\begin{enumerate}
    \item \(f^*\) is nonmonotonic with the fall off rate except for the XX model with \(\lambda =0\) (see Fig. \ref{fig:fstarwithalpha}) and  \(f^*\) is fluctuating when \(\alpha \ll 2\), irrespective of other system parameters.
    \item Interestingly, \(f^*\) reaches its maximum value when \(\alpha \approx 2\) and the maximum value, denoted as \(f^{*^{\max}}_\alpha\),  depends on the anisotropy parameter, \(\lambda\). Specifically, with \(g>1\), if one increases \(\lambda \) so that it is close to unity, \(f^*\) starts increasing when \(\alpha \gtrsim 2 \),   after achieving its maximum, it decreases and finally saturates to a certain value, \(f^{*^{sat}}_\alpha\) for \(\alpha \gtrapprox 8\).
    
    
\end{enumerate}
In other words, when the evolving Hamiltonian contains more anisotropy, the non-monotonic behavior of \(f^*\) is more pronounced, i.e.,    \(\Delta_{f^*}=f^{*^{\max}}_\alpha - f^{*^{sat}}_\alpha\) increases with the increase of \(\lambda\). For example, for the transverse Ising  model with \(\lambda =1\),  \(\Delta_{f^*} = 0.08 \) (when \(g=1.7\) and \(N=20\)) while for \(\lambda=1.3\), \(\Delta_{f^*}=0.15\) \textcolor{black}{(see Table \ref{table:1} for more details).} 
Similar nonmonotonic behavior can also be observed for other pairs of \((\lambda, g)\) and \(N \leq 100\). However, with the increase of \(N\), \(f^*\) including \(f^{*^{sat}}_\alpha\) and \(f^{*^{\max}}_\alpha\)  decreases which will be addressed next.  

\subsection{Overcoming decline of fidelity with system-size by LR interactions}
\label{sec:scaling}

The preceding investigation is carried out by fixing the length of the chain and varying other system parameters. We are now interested to study how the maximum fidelity \(f^*\) decreases with the increase of \(N\). The decaying behavior of \(f^*\) is expected as one rises the length of chain whose end points are used to place the initial and the final states. 
\begin{figure}
    \includegraphics[width=\linewidth]{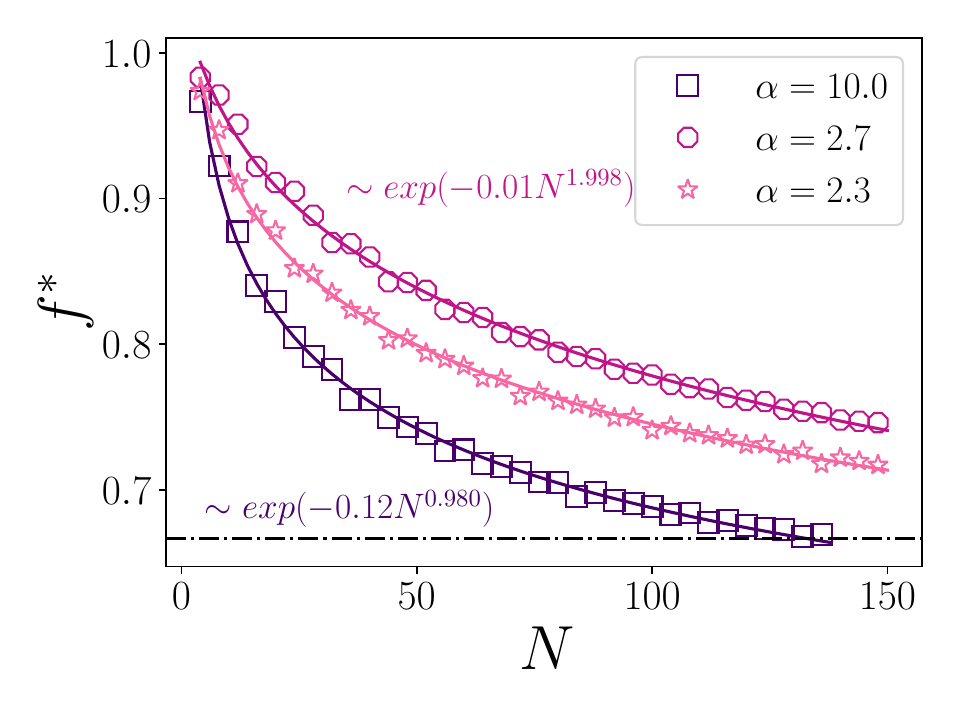}
    \caption{ The dependency of fidelity \(f^*\) $($ordinate$)$ on the system size \(N\) $($abscissa$)$ is plotted. Stars, circles, and squares represent \(\alpha=2.3\), \(2.7\) and \(10\) respectively. 
    Here $\lambda = 1$ and \(g = 1.7 \). We fit the decaying curve with \(a \exp(-b N^{\eta})\). We  find that with \(a=1\),  \(\eta\) decreases along with the increase of \(b\). Two observations emerge -- (1) for a given \(N\), \(f^{*^{\alpha =10}} < f^{*^{\alpha <10}}\); (2) there exists \(N\) for which   \(f^{*^{\alpha =10}}\) cannot beat the classical limit while \(f^{*^{\alpha \sim 2}}\) is much higher than the classical fidelity.  The coordination number is $z=N-1$. 
    All the axes are dimensionless.}
    \label{Fig:fstarwithNscaling}
\end{figure}

Since  our objective is to focus on the range of interactions, the maximum fidelity is examined with the length of the chain when the coordination number is maximum $(z=N-1)$. When we consider the XX model with nearest-neighbor interactions (\(\alpha =10\)), we observe that \(f^*\)  remains above the classical limit for a very high value of \(N\) \(> 150\), while \(f^*\) sharply decreases with \(N\) when \(\alpha \approx 2\) and it reaches the classical limit with \(N\approx \mathcal{O}(100)\). 

The opposite picture emerges when one increases \(\lambda\), close to the extended Ising Hamiltonian.  In particular, we observe that when \(\lambda =1\), \(g=1.7\) with \(\alpha = 10\), \(f^* = f^{cl}\) when \(N =138\) while \(f^*\) is surely higher than the classical fidelity with the same length of the chain, i.e., \(f^* > f^{cl}\) for \(N\geq  138\) for \(\alpha \approx 2\). 

To make the analysis more concrete, we fit the decaying curve of \(f^*\) with the function, given by \(a \exp(-b N^{-\eta})\),  for different \(\alpha\) values. In order to justify the positive impact of LR interactions towards combating the decline of fidelity with system-size, we find out  how \(\eta\) and \(b\) change with the fall-off rate \(\alpha\) for maximum coordination number. Interestingly, there are system parameters for which we notice that \(b\) increases while \(\eta\) decreases with the increase of \(\alpha\). It ensures that the rate of fall for \(f^*\) with the increase of system size, which can be delayed with the introduction of LR interactions, as depicted in Fig. \ref{Fig:fstarwithNscaling}. 

\textcolor{black}{\emph{Note.} Let us argue that the small values of \(z\), with small to moderate \(\alpha\), can also be meaningful for a finite system size as one can still tune the strength of the corresponding few-body interaction terms. In this case, the performance of the QST protocol is modified primarily by the effective strengths of these \(k\)-body interactions. Moreover, in realistic physical systems, typically only a small number of neighboring spins can interact simultaneously, implying that \(z\) is naturally limited to small values. Our results demonstrate that carefully manipulating these few-body interaction terms can indeed enhance the performance of the QST protocol.}

{\color{black}
\subsection{Correspondence between entanglement and fidelity}
\label{sec:ent_logneg}

In order to understand the physical mechanism behind the state transfer protocol described in the preceding subsection, we analyze the entanglement dynamics by introducing an extra qubit \(q\) into the system,  initially entangled maximally with the first qubit which is carrying the arbitrary state to be transferred to the qubit \(N\). We analyze the flow of entanglement, calculated between the qubit \(q\) and \(N\) via logarithmic negativity \cite{plenio_prl, vidal_pra_2002} with time and establish its relation with the fidelity obtained at the terminal site (see Appendix \ref{sec:appendixCent} for the explicit calculation). Mathematically, we compute
\begin{equation}
E(\rho_{qN}) \equiv \log_2 ||\rho_{qN}^{T_{q}}||,
\end{equation}
where \(||\rho||\) represents the trace norm and \(T_{q}\) is the partial transposition with respect to \(q\) \cite{peres_prl_1996, horodecki_pla_1996}. Interestingly, we observe that the logarithmic negativity exhibits a sharp peak at the same time when the fidelity achieves its maximum value (see Fig. \ref{fig:fident}). This supports the view that the protocol’s performance is directly linked to the transient entanglement created between the two ends of the chain. 

}

\begin{figure}
    \centering
    \includegraphics[width=\linewidth]{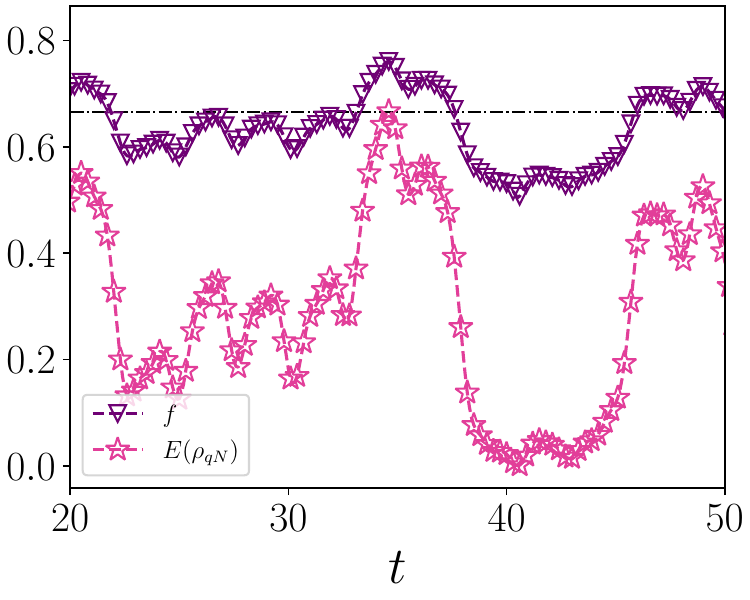}
    \caption{{\color{black}Variation of \(f\) (ordinate) and \(E(\rho_{qN})\) (ordinate) with respect to time (abscissa). Here, \(N=25\). The other parameters of the long-range Hamiltonian are \(\lambda = 0.0\), \(g = 0.7\), \(\alpha = 0.6\) and \(z=16\). Triangles and stars represent the fidelity  and the corresponding entanglement calculated via logarithmic negativity. The horizontal dashed-dot line represents the classical limit \(2/3\) of fidelity. The graph clearly shows the correspondence between the two, aligning both the maxima(s) at \(t=34.6\).  All the axes are dimensionless.}} 
    \label{fig:fident}
\end{figure}


\section{Less time to achieve quantum advantage through  long-range interactions} 
\label{sec:quantadv}

\begin{figure*}
    \includegraphics[width=0.75\linewidth]{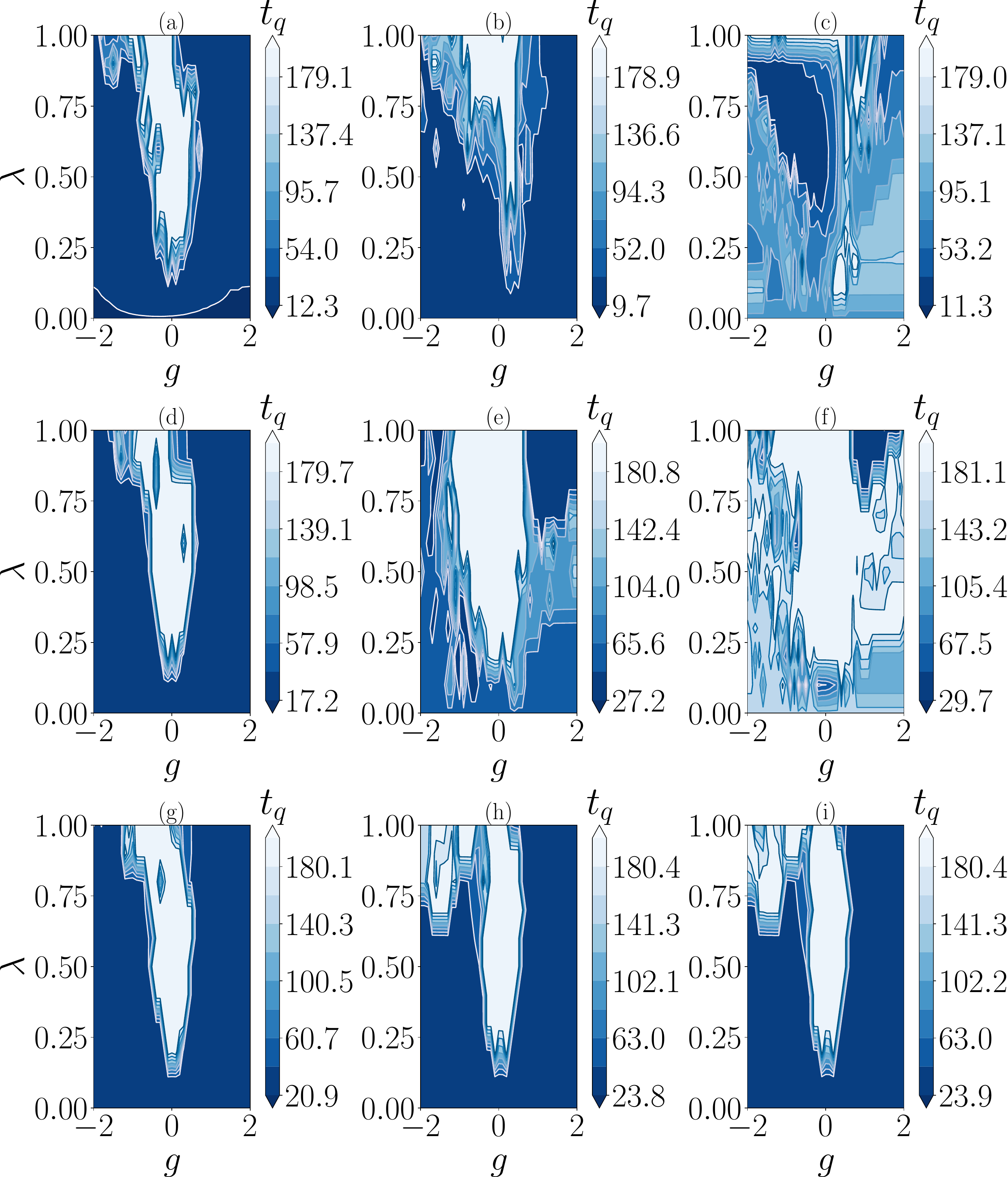}
    \caption{Contour plot of minimum time required to achieve classical fidelity, \(t_q\)  against system parameters, \(g\) (abscissa) and $\lambda$ (ordinate). Each plot corresponds to a given configuration of power-law decay rate $\alpha$ and coordination number $z$ as the following: $(a): \alpha = 0.5, z=2,(b): \alpha = 0.5, z=10,(c): \alpha = 0.5, z=N-1, (d): \alpha = 1.5, z=2, (e): \alpha = 1.5, z=10,(f): \alpha = 1.5, z=N-1, (g): \alpha = 2.5, z=2, (h): \alpha = 2.5, z=10, (i): \alpha = 2.5, z=N-1$. The color of the plots varies from dark to light, which corresponds to the smallest to the largest $t_q$. Here, the system size is $N=25$. {\color{black} Since our focus is on the parameter regimes in which \(t_q\) is small and the fidelity is high, we impose a maximum cutoff of order \(t_q \sim \mathcal{O}(10^2)\). All values above this cutoff are considered equivalently large and are displayed in white, which serves to highlight the regions of interest more clearly. } All axes are dimensionless. } 
    \label{fig:all_z_alpha_t_q}
\end{figure*}

{\color{black}After guaranteeing the benefit of long-range interacting Hamiltonian as evolving one with respect to the maximum achievable fidelity, it is  now natural to ask -- ``{\it what is the minimum  time required (\(t_q\)) to transfer a quantum state from one end of the spin chain (or quantum channel) to the other, while surpassing the classical fidelity limit?"} 
 To address this, we determine the minimum time at which the average fidelity goes beyond \(2/3\) for a fixed length of a spin chain, thus ensuring a {\it quantum advantage}. The most favorable scenario is that the minimum transfer time is small while the maximum achievable fidelity is high. 
 
In order to find \(t_q\), we compute the lowest time when 
 \[
 f-\frac{2}{3} > \epsilon,
 \] 
 where \(\epsilon\) is a small positive tolerance number, in our analysis, \(\epsilon \sim 10^{-4}\). This figure of merit can be attributed to the basic necessity for transferring quantum states.  We will analyze how \(t_q\) depends on the variable-range interactions involved in dynamics and emphasize that 
 it also depends crucially on the length of the chain \(N\) as we are interested in studying the state transfer protocol. }
 
\begin{figure*}
    \centering
    \includegraphics[width=\linewidth]{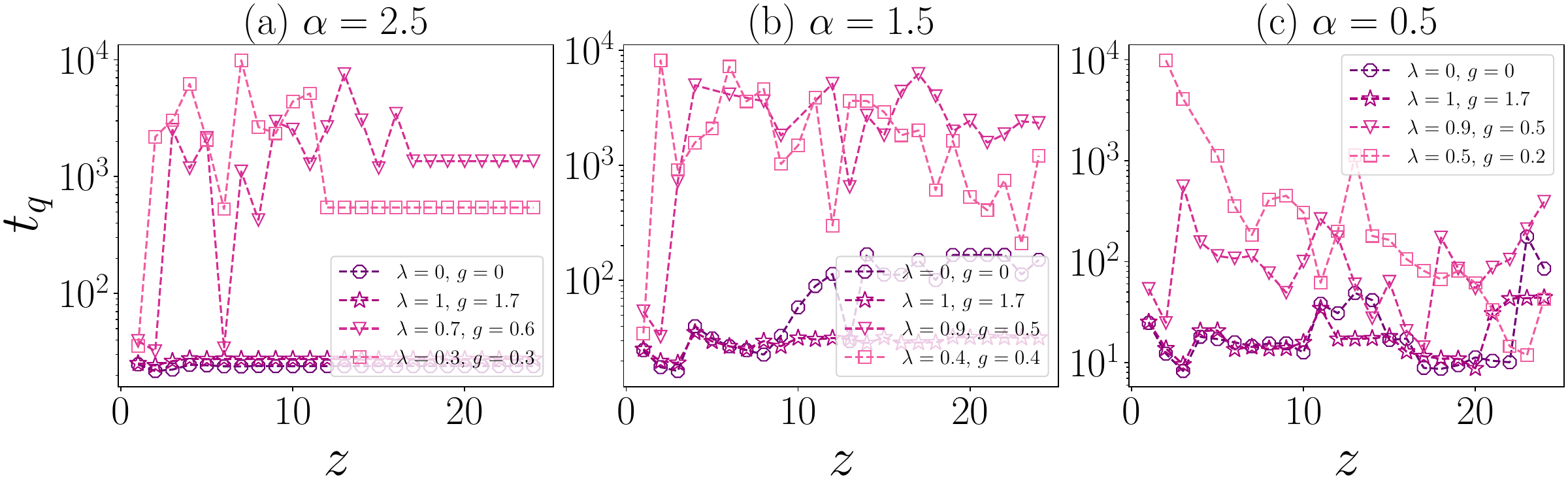}
    \caption{Illustration of  \(t_q\) $($ordinate$)$  versus coordination number \(z\) $($abscissa$)$ for the extended transverse $XY$ model. To demonstrate, we choose  four pairs of \( (\lambda, g)\) depicted in varying color shades, corresponding to three regimes of $\alpha$ ( i.e., $2.5, 1.5$, and $0.5 $), belonging to short-, quasi long- and long-range regimes. Here \(N = 25\). The results highlight the preference for short- and long-range interactions across different regimes depending upon the choice of \( (\lambda, g)\) values. In both quasi long- and long-range domains, there always exists \(z>1\) values for which \(t_q\) is smaller than that obtained with \(z=1\), corresponding to NN interactions. All the axes are dimensionless. } 
    \label{Fig:tqzvariation}
\end{figure*}
\subsection{Crucial role of system parameters in quantum advantage}
\label{sec:systempara}


{\color{black}
Let us first concentrate on moderate values of \(z (\sim N/2)\). In this case, two observations emerge. First, when \(\alpha \in [2, 3]\), there exists a region in the \((\lambda,g)\)-plane which provides low \(t_q\) values, independent of \(z\). Second, for \(\alpha<2\), parameter regimes with  \(|g|>1\) emerge as optimal for minimizing the time required to achieve quantum advantage, largely independent of the anisotropy parameters 
 (see the second column of Fig. \ref{fig:all_z_alpha_t_q}).  Moreover, these trends clearly distinguish the regimes 
\(\alpha >2\) and \(\alpha \leq 2\), emphasizing the different roles played by long-range interactions determining \(t_q\).

In the extreme connectivity limit,  \(z = N-1\), although  the choices of \(\lambda\) and \(g\) values do not vary much as observed for \(z\sim N/2\), \(t_q\) here fluctuates with \(\alpha\), thereby exhibiting pronounced sensitivity to the fall-off rate \(\alpha\).  Specifically, \(t_q\) increases sharply when \(\alpha <2\) while  for \(\alpha >2\), it remains essentially unchanged  for all values of \(z\) (see the third column of Fig. \ref{fig:all_z_alpha_t_q}). This behavior can be attributed to the properties of the ground state entanglement  in variable-range interacting systems. In particular, 
in the quasi long-range regimes \((1 < \alpha\leq 2)\), the entanglement length is comparable to that of  short-range interacting systems while  when \(\alpha <1\), it exhibits a qualitatively different scaling 
behavior~\cite{GaneshDebasisLR}. 

We now turn to another limiting case, namely interactions just beyond the nearest-neighbor regime, corresponding to \(z=2\), which nevertheless incorporate few-body interaction effects. When \(z=2\),  the values in the \((\lambda,g)\)-plane, which are favorable for minimizing \(t_q\) with \(\alpha>2\) remain so even for low \(\alpha\) values.   For example, when \(z=2\) and \(0\leq \alpha\leq 2\), \(t_q \leq 30\)  for all values of \(\lambda\) and \(|g|\)  except few regions when \(g \sim 0\) for all \(\lambda\) and \(g<-1\) with \(\lambda \sim 1\) (see the first column of Fig. \ref{fig:all_z_alpha_t_q}). 

It is also important to note here that the analysis is performed with \(N \leq 50\) and can be the artifact of finite-size systems. It is obvious that if one increases \(z\) values, \(t_q\) starts depending on \(\alpha\) more crucially which is not the case for low \(z\) values, e.g., \(z=2\). 

 }

\begin{figure}
    \centering
    \includegraphics[width=\linewidth]{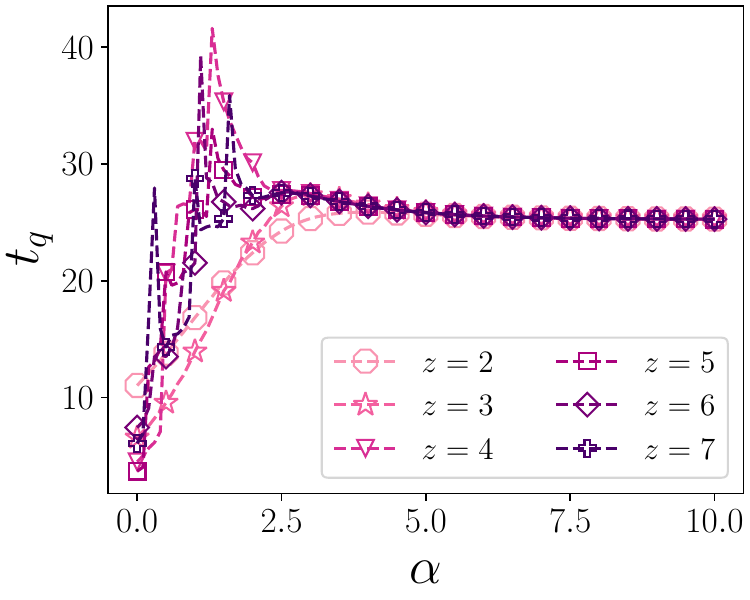}
    \caption{The variation of time \(t_q\) $($ordinate$)$ with respect to the decay strength $\alpha$ $($abscissa$)$.  Increasing coordination numbers \(z\)  represented by a gradient from light to dark shades (with different markers), is presented, corresponding to \(N = 25\), with \( \lambda = 1\) and  \(g = 1.7\). We observe that when \(z\ \leq N/2\), \(t_q\) required for low \(\alpha\) is less than the one with high \(\alpha\) although there are fluctuations in \(t_q\) with low \(\alpha\). \(t_q\) saturates when \(\alpha \geq 4\) \(\forall z\leq N/2\).  All the axes are dimensionless. } 
    \label{fig:tqwithalphadiffz}
\end{figure}

\begin{table}[h!]
\centering
\begin{ruledtabular}
\begin{tabular}{ccccc}
$\lambda$ & $g$ & $\alpha$ & $z$ & $t_q$ \\
\hline
1   & 1.7 & 1.2 & 2  & 24.86 \\
0.4 & 0.4 & 1.5 & 2  & 28.91 \\
1   & 1.7 & 0.4 & 4  & 18.45 \\
0   & 0   & 3.6 & 7  & 27.73 \\
0.9 & 0.5 & 0.5 & 10 & 46.86 \\
0.3 & 0.3 & 2.5 & 24 & 36.46 \\
\end{tabular}
\end{ruledtabular}
 \caption{{\color{black}Variation of the minimum time \(t_q\) for different system parameters. Here, \(N=25\). It represents the fast and better state transfer in a spin chain with the help of long-range interactions strength \(\alpha\) and \(z\).}}
\label{table:2}
\end{table}

\subsection{Improvement of minimum time required to attain quantum advantage with coordination number}
\label{sec:coordinationno}

{\color{black} The above analysis allows us to identify an appropriate set of parameters, \(g\) and \(\lambda\), that  minimize \(t_q\). Fixing the pair \( (\lambda, g) \) and \(N\), we  now investigate how LR interactions influence quantum state transfer by systematically varying the coordination number \(z\) and interaction strength, \(\alpha\).  From the above discussion, it is  evident  that the quantum advantage can be obtained when both short- and long-range interacting Hamiltonian are used for dynamics. However, our goal here is to establish an ordering among the characteristic times,  \(t_q^{0\leq \alpha \leq 1}\), \(t_q^{1< \alpha\leq 2}\) and \(t_q^{2 < \alpha \leq 3}\) where the superscripts denote the \(\alpha\) values belonging to long-, quasi long- and short-range interaction regimes \textcolor{black}{(according to thermodynamic limit). }Furthermore, it is important to emphasize that, since our analysis is restricted to finite system sizes, the short-range regime defined in the thermodynamic limit may not hold here, and residual effects of long-range and few-body interactions can therefore persist.


{\it Short-range but close to quasi-LR domain.}
We first consider the short-range regime with interaction strength close to the quasi-LR boundary. 
For  fixed \(2\leq \alpha<3\), and for all values of \(g\) and \(\lambda\), the dependence of \(t_q\) on the coordination number \(z\) closely mirrors the behavior of the maximum fidelity,
\(f^*\) (comparing Figs.  \ref{fig:fstartwithzdiffalpha} and \ref{Fig:tqzvariation}). Specifically,   \(t_q\)  initially decreases (increases) with \(z\) and then increases (decreases) before saturating to a fixed value,  \(t_q^{sat}\), at large \(z\). Again, comparing the result with \(z=1\),  we observe that
when \(|g|<1\),
\begin{eqnarray}
t_q^{z=2} < t_q^{z=1}, \, \, \text{and}\, \,   t_q^{sat} < t_q^{z=1},   
\end{eqnarray} 
independent of the anisotropy parameters, provided \(\lambda \neq 0\). For example, \(\lambda=0.5\),  \(g=0.7\) and \(\alpha =2.3\), we find \(t_q^{z=1} = 30.84 > t_q^{z=2} =25.23\) and   \(t_q^{sat} = 28.7\).

However, one of the above inequalities is violated for \(|g|>1\) for different anisotropy parameters. 
Overall, there exist  pairs of \((g, \lambda)\) for which \(t_q\) remains nearly independent of
 \(z\) while for some values, it exhibits noticable fluctuations  with \(z\). On average, we find \(t_q \approx [20, 30] \) in this domain of \(\alpha\). 

{\it Quasi-LR regime.} Again, \(|g|>1\), \(t_q\)  remains almost constant with \(z\) for some \(\lambda\)  values including \(\lambda =0\). Interestingly, in this domain,  one can always identify a coordination number \(z>1\)  for which 
\[
t_q^{z=1} > t_q^{z>1},
\] 
as illustrated in Fig. \ref{Fig:tqzvariation}(b)). On the other hand, when \(|g|<1\), and \(\lambda \neq 0\), \(t_q^{z>1}\) fluctuates more and is significantly higher than \(t_q^{z=1}\).  While this increase can be attributed to enhanced entanglement creation between distant sites which is not the case when there is only nearest-neighbor interactions. However,  we will show that such a simplistic explanation may not be true since low \(t_q\) can also be found in the LR domain. 

{\it \(0 \leq \alpha \leq 1\) -- LR domain.} 
Turning to the genuinely LR interactions, we observe behavior qualitatively similar to that found in the short- and quasi long-range models. In particular, 
 there exist tunable \((\lambda, g)\)-pairs for which  certain values of \(z\) satisfy \(t_q^{z=1} > t_q^{z>1}\). In these cases, the variation of \(t_q\) with \(z\) is not substantial. To capture this,  we compute the average \(t_q\) over all \(z\) for a fixed \(\alpha\) and \(N\), which we denote as \(\overline{t_q}\). For example,  for \(N=25\),  \(\lambda =0\) and \(\forall g\), we find that for \(\alpha =2.5\),  \(\overline{t_q} = 23.82\), and for \(\alpha =1.5\),  \(\overline{t_q} = 87.48\) while when we use  LR interacting Hamiltonian for state transfer, we obtain  \(\overline{t_q} = 27.89\) with \(\alpha=0.5\) (see Fig. \ref{Fig:tqzvariation}) . At the same time, we also identify the system parameters, for which \(t_q^{z=1} > t_q^{z>1}\, \forall z\), and the difference is substantial. For example,  with \(\lambda =0.5, g=0.2\) and \(N=25\), we notice that \(t_q^{z=1} \sim \mathcal{O}(10^4)\) and  \(t_q^{z=24} = 42.42\) (see Table \ref{table:2} to obtain a clear picture). This, once again, clearly establishes the usefulness of  LR interacting spin model for the state transfer scheme which naturally occurs in several physical systems.

We also find anisotropy and magnetic field strengths for which \(t_q \sim \mathcal{O}(10^3)\), i.e., the minimum time required to achieve quantum advantage is comparatively high to the one obtained for \(\lambda =0\) and other \(\lambda\) values with \(g>1\). Interestingly, for certain tunable parameter pairs   \((\lambda, g)\), \(t_q\) decreases with increasing \(z\), which again  highlights  quantum advantage enabling by LR interactions (see Fig. \ref{Fig:tqzvariation} (c)).  Overall,  these observations reveal that the fall-off rate and the range of interactions influence state transfer through a subtle interplay between entanglement creation and destruction. 
}


\subsubsection{ Better state transfer with low coordination number}

From the study, it is clear that \(z \leq N/2\) has a special status although they involve interactions beyond nearest neighboring sites. In this situation, \(t_q\)  increases with \(\alpha\) and saturates to a certain \(t_q^{sat}\) value irrespective of the anisotropy and magnetic field strength provided the length of the spin chain is moderate (see Fig. \ref{fig:tqwithalphadiffz}). The saturation  occurs when the fall off rate, \(\alpha\), belongs to the short-range regimes. Precisely,  the saturation of time for quantum advantage, \(t_q^{sat}\), happens when \(\alpha \gtrapprox 4\) for a fixed \(N \sim \mathcal{O}(50)\). This again indicates that when \(\alpha\) is less than \(4\), we can surely gain by employing long-range interactions during the dynamics. When \(z\) is relatively high and close to \(N/2\), some fluctuations in \(t_q\) are also observed for \(\alpha \leq 2\) although \(t_q^{\alpha =0} < t_q^{sat}\). This indicates that although there can be some advantages of long-range interacting systems, it becomes highly sensitive towards a successful state transfer, especially in the quasi long-range regimes as is evident from \(\overline{t_q}\). As discussed earlier, the competition between two kinds of operations, entanglement-generating and disentangling power \cite{disenta_power_sibasish_2007} of the evolution operator, can be responsible for fluctuations in \(t_q\) when the fall-off rate is small or moderate for which the unitary operator has a high multipartite entanglement-generating capability \cite{ent_power_zanardi_2001,gorshkov_qst_3} compared to the system having only nearest-neighbor interactions. 


\section{Conclusion}
\label{sec:conclu}

The extended $XY$ model with long-range interactions provides a powerful platform for investigating the feasibility of quantum state transfer (QST) protocols and exploring their potential in quantum technologies. We demonstrated that the long-range interactions had a direct impact on enhancing the maximum fidelity of state transfer, even in larger systems. While fidelity typically decreased with an increase in system size, the presence of long-range interactions reduced this decline, ensuring reliable performance in scenarios where short-range models failed to maintain fidelity above the classical threshold. It was also found that the interplay of anisotropy and magnetic field strength played a crucial role, with strong anisotropy enhancing fidelity at the expense of increased sensitivity to field variations, while the XX model, with a vanishing anisotropy limit, demonstrated robustness over a wider range of conditions. {\color{black} We also established the physical mechanism behind the long-range benefit via the entanglement generation between the first and the last qubit.}

The study revealed that incorporation of long-range interactions in the spin chain significantly altered the dynamics of state transfer, resulting in a shorter time required to achieve quantum advantage. Specifically, it was observed that the quasi-long- and long-range regimes, where the decay strength is moderately low, exhibited an optimal interplay between interaction range and efficiency, outperforming both short-range systems. Furthermore, the role of the coordination number in QST was highlighted, where intermediate coordination numbers, involving interactions beyond nearest neighbors, facilitated faster and more reliable state transfer compared to excessively high values. 

Additionally, distinct behaviors across the long-, quasi long-, and short-range interaction regimes were uncovered. In the long-range regime, fluctuations in the fidelity and transfer time were attributed to the dynamics of entanglement generating and disentangling powers of the evolving Hamiltonian, whereas  the quasi long-range domain provided a stable and efficient configuration for state transfer. In contrast, the short-range interacting model closely resembled traditional nearest-neighbor ones, exhibiting diminished performance due to limited interaction range.

These findings are not only of theoretical interest but they can also be experimentally verified in state-of-the-art quantum platforms, including trapped ions, Rydberg atom arrays, and optical lattices. The parameter regimes explored in this work align well with the capabilities of these systems, where decay strength, coordination number, anisotropy, and magnetic field strength can be finely tuned. By leveraging the exotic properties of long-range interactions, this study has opened potential avenues for the design of scalable and efficient quantum communication systems,  either on their own or within quantum computing circuits,  providing a solid foundation for future advancements in the field.

\acknowledgments

We  acknowledge the use of cluster computing facility at the Harish-Chandra Research Institute. This research was supported in part by the INFOSYS scholarship for senior students. We acknowledge support from the project entitled ``Technology Vertical - Quantum Communication'' under the National Quantum Mission of the Department of Science and Technology (DST)  (Sanction Order No. DST/QTC/NQM/QComm/$2024/2$ (G)).
 LGCL received funds from project DYNAMITE QUANTERA2-00056 funded by the Ministry of University and Research through the ERANET COFUND QuantERA II – 2021 call and co-funded by the European Union (H2020, GA No 101017733).  Funded by the European Union. Views and opinions expressed are however those of the author(s) only and do not necessarily reflect those of the European Union or the European Commission. Neither the European Union nor the granting authority can be held responsible for them. This work was supported by the Provincia Autonoma di Trento, and Q@TN, the joint lab between University of Trento, FBK—Fondazione Bruno Kessler, INFN—National Institute for Nuclear Physics, and CNR—National Research Council. TKK acknowledges
the support from the second Swiss Contribution MAPS (Grant No. 230870).

\appendix

\section {Diagonalization of $XY$ model}
\label{sec:appendixA}

Let us begin by considering a general quadratic Hamiltonian \cite{lieb1961} in fermionic operators:
\begin{equation}
\hat{\mathcal{H}} = \sum_{i,j} P_{ij} \hat{f}_i^\dagger \hat{f}_j + \frac{1}{2} \sum_{i,j} (Q_{ij} \hat{f}_i^\dagger \hat{f}_j^\dagger + Q_{ij}^* \hat{f}_j \hat{f}_i)
\end{equation}
Here, $\hat{f}_i$ and $\hat{f}_i^\dagger$ are fermionic annihilation and creation operators, respectively,  obeying the canonical anticommutation relations, $\{\hat{f}_i, \hat{f}_j^\dagger\} = \delta_{ij}$ and $\{\hat{f}_i, \hat{f}_j\} = \{\hat{f}_i^\dagger, \hat{f}_j^\dagger\} = 0$. The matrices $P$ and $Q$ encode the specific properties of our system, including interaction strengths and coupling terms. To diagonalize this Hamiltonian, we introduce Bogoliubov quasiparticle operators as
\(\hat{\eta}_q = \sum_{m} (\mathcal{A}_{qm} \hat{f}_m + \mathcal{B}_{qm}\hat{f}_{m}^\dagger)\). 
The transformation matrices $\mathcal{A}$ and $\mathcal{B}$ are chosen such that the Hamiltonian becomes diagonal in the new basis $\hat{\mathcal{H}} = \sum_{q} \epsilon_q \hat{\eta}_q^\dagger \hat{\eta}_{q}$, where $\epsilon_q$ represents the quasiparticle energies. This diagonalization is crucial for understanding the system's excitation spectrum and its response to perturbations.

Now, let us connect this general form to our specific spin system. We can map our spin-$1/2$ operators to fermionic operators using the Jordan-Wigner transformation. This well-known transformation can be expressed as $\hat{S}_{j}^{+} = \hat{f}_{j}^\dagger \prod_{i<j} (1 - 2\hat{f}_i^\dagger \hat{f}_{i}), 
\hat{S}_{j}^{-} = \hat{f}_{j} \prod_{i<j} (1 - 2\hat{f}_i^\dagger \hat{f}_{i}),
\hat{S}_{j}^{z} = \hat{f}_j^\dagger \hat{f}_{j} - \frac{1}{2}
$, where $\hat{S}_{j}^{\pm} = \hat{S}_{j}^{x} \pm i\hat{S}_{j}^{y}$ are the spin raising and lowering operators.  The string operator $\prod_{i<j} (1 - 2\hat{f}_i^\dagger \hat{f}_{i})$ ensures the correct anticommutation relations between operators at different sites, maintaining the non-local nature of the spin algebra in the fermionic representation. Applying this transformation to our spin Hamiltonian yields Eq. (\ref{eq:LRHam}) \cite{barouch_pra_1970_1,lieb1961,barouch_pra_1970_2}.

To study the dynamics, we work in the Heisenberg picture. The time evolution of the fermionic operators for a given site \(m\) is given by
\begin{equation}
\hat{f}_m(t) = e^{i\hat{\mathcal{H}}t}\hat{f}_{m}e^{-i\hat{\mathcal{H}}t} = \sum_{k=0}^N [\Phi_{mk}(t)\hat{f}_k + \Psi_{mk}(t)\hat{f}_{k}^\dagger], 
\end{equation}
Here the time-dependent coefficients $\Phi(t)$ and $\Psi(t)$ are determined by the Bogoliubov transformation, 
\(\Phi(t) = \mathcal{A}^T e^{-i\epsilon t}\mathcal{A} + \mathcal{B}^T e^{i\epsilon t}\mathcal{B} \) and 
\(\Psi(t) = \mathcal{A}^T e^{-i\epsilon t}\mathcal{B} + \mathcal{B}^T e^{i\epsilon t}\mathcal{A}\)
These equations provide a complete description of the system's time evolution, allowing us to calculate various observables and study both equilibrium and non-equilibrium properties of the system.

\section{ Computation of fidelity}
\label{sec:appendixBfid}

This Appendix outlines the derivation of fidelity $f$ for quantum information transfer in systems with long-range interactions, building upon the framework presented in Ref. \cite{Bayat11}. Let $\Lambda$ be a quantum channel that describes the evolution of the density matrix from the  qubit at \(t = 0\) to the resulting qubit  at time \(t\) as $\sigma(t) = \Lambda[\sigma(0)]$. The fidelity $f$ which characterizes the channel's efficiency is defined as \cite{horodeckiPRA} 
$
\Omega(t) = \max_{V \in U(2)} \int d\phi_i \bra{\phi_i} V^{\dagger} \Lambda[\ket{\phi_i}\bra{\phi_i}] V \ket{\phi_i},
$ where V is the local unitary operation. For systems governed by anisotropic exchange interactions, assuming conservation of excitation number and focusing on the Bloch sphere mapping from the initial to the resulting qubit such that the corresponding fidelity can be expressed as
\(f = \frac{1}{2} + \frac{1}{6}\big|p(t)^2 - q(t)^2\big| + \frac{1}{3}\max\{p(t), q(t)\}\),
where $p(t) = |\Phi_{N0}(t)|$ and $q(t) = |\Psi_{N0}(t)|$ are time-dependent coefficients related to the transfer process. The coefficients $p(t)$ and $q(t)$ are determined by the system's Hamiltonian parameters and incorporate the effects of long-range interactions. 

{\color{black}
\section{Computation of entanglement}
\label{sec:appendixCent}

The qubit \(q\) is entangled with the first qubit of our protocol via Bell state \(|\Phi^+\rangle\).
Using the superoperator \(\mathcal{E}_t\), which describes the map from the initial to the final state, the bipartite state of qubits \(q\) and \(N\) can be expressed as \cite{Bayat11}
\begin{equation}
\rho_{qN} (t) = (I\otimes \mathcal{E}_t)[|\Phi^+\rangle \langle \Phi^+|].
\end{equation}
The superoperator can be constructed from
\begin{equation}
\langle i| \rho_{N} (t)|j \rangle = \sum_{l,m = 0}^1 \langle i | \mathcal{E}_t(|l\rangle \langle m|)|j \rangle \langle l|\rho_1 (0)|m \rangle.
\end{equation}
The corresponding choi matrix can be obtained as
\begin{equation}
\langle l i| \mathcal{C}_{\mathcal{E}} (t)| mj \rangle = \langle i | \mathcal{E}_t(|l\rangle \langle m|)|j \rangle,
\end{equation}
with
\begin{widetext}
\begin{equation}
    \mathcal{C}_{\mathcal{E}}(t)=\left[
\begin{array}{cccc}
  |\mathcal{A}_{N1} (t)|^2 + 1 & 0 & 0 & |\mathcal{A}_{N1} (t)|e^{i\theta_a(t)} \\
 0 & -|\mathcal{A}_{N1}(t)|^2 & |\mathcal{B}_{N1} (t)|e^{i\theta_b(t)} & 0 \\
 0 &  |\mathcal{B}_{N1} (t)|e^{-i\theta_b(t)} & |\mathcal{B}_{N1}(t)|^2 + 1 & 0 \\
  |\mathcal{A}_{N1} (t)|e^{- i\theta_a(t)} & 0 & 0 & - |\mathcal{B}_{N1}(t)|^2 \\
\end{array}
\right ].
\end{equation}
\end{widetext}
Here \(\theta_a (t) = arg[\mathcal{A}_{N1} (t)]\) and \(\theta_b (t) = \pi - arg[\mathcal{B}_{N1} (t)]\).
Using the results in Refs. \cite{horodeckiPRA,horo_2000}, the density matrix representing the bipartite state of \(q\) and \(N\) reads \(\rho_{qN} = \frac{1}{2} \mathcal{C_{E}} (t)\). 

}

\bibliography{reference}

\end{document}